\documentclass[fleqn,12pt,twoside]{article}
\usepackage{espcrc1}

\usepackage{graphicx}
\usepackage[figuresright]{rotating}

\newcommand{\AmS}{{\protect\the\textfont2
  A\kern-.1667em\lower.5ex\hbox{M}\kern-.125emS}}

\title{Efficient Searches for $r$-Process-Enhanced, Metal-Poor Stars}

\author{T.C. Beers, \address{Department of Physics \& Astronomy and JINA: Joint Institute for
        Nuclear Astrophysics, Michigan State University, USA} 
        P.S. Barklem, \address{Department of Astronomy \& Space Physics, Uppsala University, Sweden}
        N. Christlieb, \address{Hamburger Sternwarte, Universit\"at Hamburg, Germany} 
        and
        V. Hill \address{GEPI, Observatorie de Paris Meudon, France}}
       
\begin{document}


\maketitle

\begin{abstract}

Neutron-capture-enhanced, metal-poor stars are of central importance to
developing an understanding of the operation of the $r-$process in the early
Galaxy, thought to be responsible for the formation of roughly half of all
elements beyond the iron peak. A handful of neutron-capture-rich, metal-poor
stars with [Fe/H] $< -2.0$ have already been identified, including the well
known $r-$process-enhanced stars CS~22892-052 and CS~31082-001. However, many
questions of fundamental interest can only be addressed with the assemblage of a
much larger sample of such stars, so that general properties can be
distinguished.

We describe a new effort, HERES: The Hamburg/ESO R-Process-Enhanced Star survey,
nearing completion, which will identify on the order of 5--10 additional highly
$r$-process-enhanced, metal-poor stars, and in all likelihood, a similar or
greater number of mildly $r$-process-enhanced, metal-poor stars in the halo of
the Galaxy. HERES is based on rapid ``snapshot'' spectra of over 350 candidate
halo giants with [Fe/H]$ < -2.0$, obtained at moderately high resolution, and
with moderate signal-to-noise ratios, using the UVES spectrograph on the
European VLT 8m telescope. 

\end{abstract}

\section{Introduction}

Over the course of the past decade, a very rare sub-class of metal-poor stars in
the halo of the Galaxy has been identified that is of great interest for
developing detailed understanding of the astrophysical rapid neutron-capture
process. These stars exhibit enhancements in their observed ratios of
$r$-process elements, relative to iron, of from 2 to over 50 times the solar
values, {\it i.e.}, $+0.3 < [r{\rm -process}/{\rm Fe}] < +1.7$. The first star
discovered in this class, CS~22892-052, a giant with [Fe/H] $= -3.1$ \cite{Sne94},
has been intensely studied in the optical with the world's largest
telescopes, as well as in the near ultraviolet with the Hubble Space Telescope
\cite{Sne03}. This star allowed the first application of the Th/Eu
cosmo-chronometer to an extremely metal-poor star, and has been used to place
direct age limits on the star, and in turn on the Galaxy and the Universe.
The second example discovered, CS~31082-001, permitted the detection of U,
and application of a new, and in principle better understood chronometer, U/Th
\cite{Cay01,Hil02}. Both of these stars, and other mildly $r$-process-enhanced
stars discovered since, share the remarkable property that their ``heavy''
$r$-process elements, i.e., in the range $56 < Z < 76$, exhibit a pattern of
$r$-process abundances that is identical to the solar $r$-process pattern, with
clear implications for the nature of the $r$-process, and perhaps on its
astrophysical site.

Many questions arise, however, as to the nature of mildly- to highly-
$r$-process-enhanced, metal-poor stars, such as (a) What is the frequency of the
enhancement phenomenon as a function of [Fe/H] ? (b) What is the distribution
of the level of enhancement -- {\it e.g.}, is it bi-modal or continuous ? (c)
With what precision is the solar $r$-process pattern reproduced from star to
star ? and (d) How common is the so-called ``actinide boost'' problem (where the
measured abundance of Th appears up to 0.4 dex {\it higher} than expected) noted
by \cite{Hil02} and \cite{Hon04} ? The answers to these, and many other
questions, requires a much larger sample of $r$-process-enhanced, metal-poor
stars to be assembled.  

\section{The HERES Approach}

The central difficulty in obtaining large samples of $r$-process-enhanced MP
stars is their extreme rarity. Based on high-resolution spectroscopy that has
been performed on the most metal-deficient giants, it appears that $r$-II stars
(for convenience, a class defined by $+1.0 < [r{\rm -process}/{\rm Fe}]$) occur
no more frequently than about 1 in 30 for giants with [Fe/H] $< -2.5$, {\it
i.e.}, roughly 3--4\%. Due to the steep decrease of the metallicity distribution
function of the Galactic halo at low metallicities, the candidates to be
examined are rare themselves, although modern spectroscopic wide-angle efforts,
such as the HK survey of Beers and collaborators, and the Hamburg/ESO survey (HES) of
Christlieb and collaborators, have succeeded in identifying such stars with
success rates as high as 10--20\%. So, we are faced with the daunting prospect
of searching for a rare phenomenon amongst rare objects. The $r$-I stars (a
class defined by $+0.3 < [r{\rm -process}/{\rm Fe}] \le +1.0$) appear to be found
with a frequency that is at least a factor of two greater than this, and
fortunately extend into the higher metallicity stars ({\it e.g.}, BD+17:3248
with [Fe/H] = $-2.1$), where a greater number of candidates exist. 

Detection of uranium presents an even bigger challenge, due to the weakness of
the absorption lines involved, and blending with features of other species. It
was not possible to measure even the strongest uranium line in the optical, U~II
3859.57 \AA\ , in the carbon­-enhanced star CS~22892--052, due to blending with a
CN line. Other lines (such as from Fe) can cause potential problems as well. The
ideal star for detecting uranium would therefore be a cool giant with low carbon
abundance, very low overall metallicity, but strong enhancement of the
$r$-process elements. It would ideally be a bright star, because high
signal-to-noise ($S/N$) ratios as well as high spectral resolving power ($R =
\lambda / \delta\lambda > 60,000$) are required to measure the strength of the U~II 3859.57 
\AA\  line accurately. Note that in CS~31082--001, this line has an equivalent width
of only a few m\AA .

Details of the techniques used to assemble the HERES sample are described by
Christlieb et al. \cite{Chr04}; here we provide a brief overview.

HERES adopts a two-step approach to the identification of neutron-capture-rich
metal-poor stars. The first step consists of the identification of a large
sample of metal­ poor giants with [Fe/H] $< -2.5$ in the HES, by means of
moderate-resolution (2 \AA ) follow-up spectroscopy of several thousand cool
($0.5 < B -V < 1.2$) metal-poor candidates selected in that survey. In the
second step, ``snapshot'' spectra ($S/N > 20$ per pixel at 4100 \AA\ ; $R \sim
20,000$) of confirmed metal-poor stars are obtained. Such spectra can be secured
for a B = 15.0 star with a 8m-class telescope in exposure times of only 15
minutes, and under less than optimal observing conditions. The weak constraints
on the observing conditions makes it feasible to observe large samples of stars.

The snapshot high-resolution spectra allow one to easily identify stars with
enhancements of $r$-process elements, using the Eu II 4129.73 \AA\  line, since
this line is very strong in these stars. For example, in CS~22892--052, it has
an equivalent width of more than 100 m\AA\ .  HERES also identifies
stars that are enhanced in $s$-process elements; these are generally
distinguishable from the $r$-process-enhanced stars by the ratio [Ba/Eu], which
is high for the $s-$process, but low in the case of the $r-$process.

We are executing the snapshot approach in a Large Programme (P.I. Christlieb)
approved by ESO. A total of 376 stars (including 4 comparison stars) are
scheduled to be observed; most of them are from the HES. These observations are
expected to yield 5--10 new $r$-II stars , and about twice as many $r$-I stars.
As a by­product, our program will provide the opportunity to measure abundances
of alpha-elements such as Mg, Ca, and Ti, and of iron-peak elements such as Cr,
Mn, Fe, Co, Ni, and Zn, as well as others, depending on the $S/N$ of each
spectrum, for the entire set of stars that we plan to observe in snapshot mode. 

Given the large number of spectra to be processed, it is mandatory that we
employ automated techniques for abundance analysis. These techniques are
described in detail in Barklem et al. \cite{Bar04}. The completion of HERES
will result in, by far, the largest sample of very metal-poor stars with
homogenously-measured abundances of a significant number of individual elements.
Figures 1 and 2 show examples of trends for Mg and Eu.  

\begin{figure}[htb]
\begin{minipage}[t]{75mm}
\includegraphics[width=7cm, height=4cm]{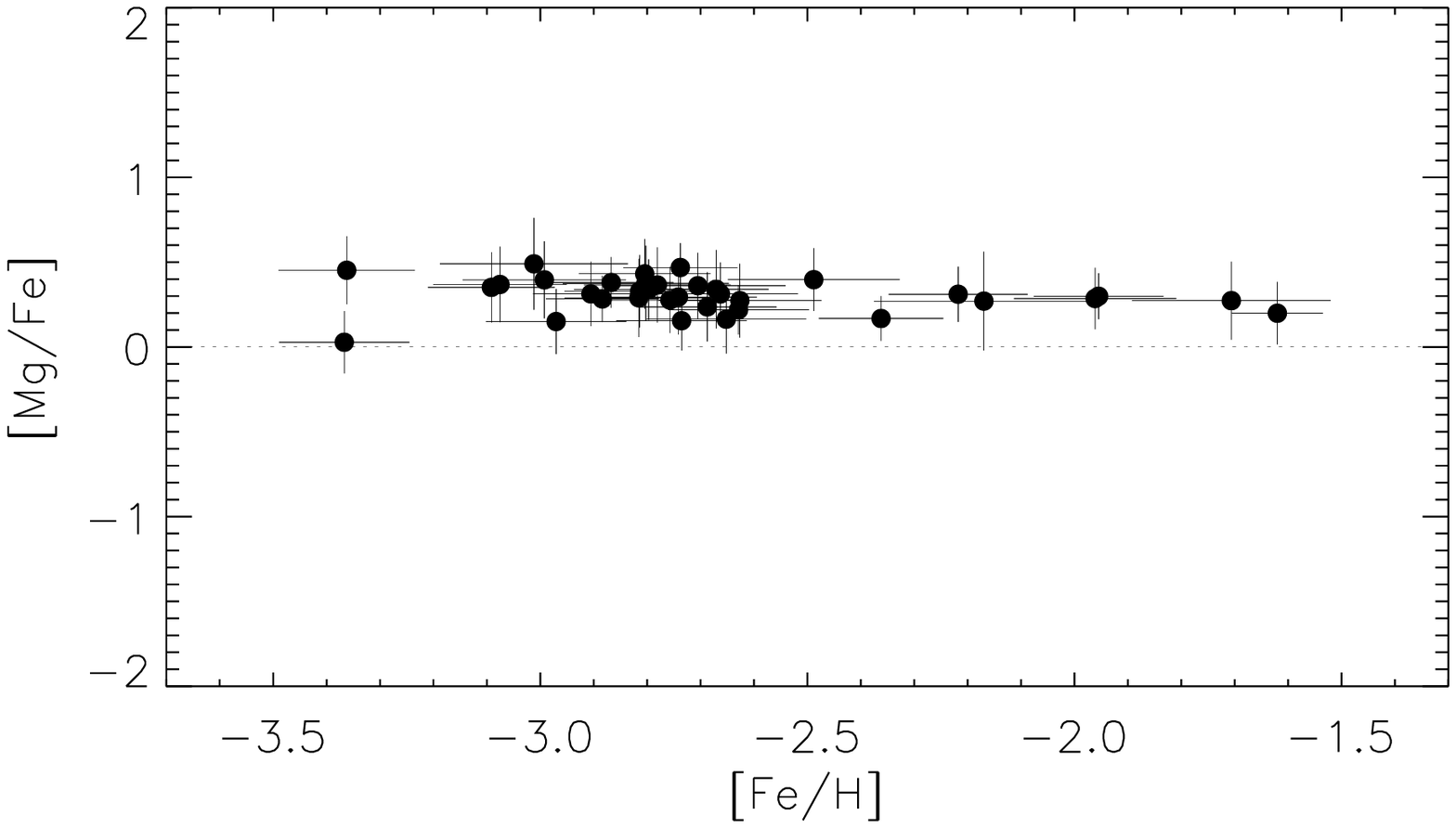}
\caption{The [Mg/Fe] vs. [Fe/H] trend exhibited for stars in the HERES Pilot Program.
When the complete HERES project is finished, the sample
will contain {\it ten times} as many stars.}
\end{minipage}
\hspace{\fill}
\begin{minipage}[t]{70mm}
\includegraphics[width=7cm, height=4cm]{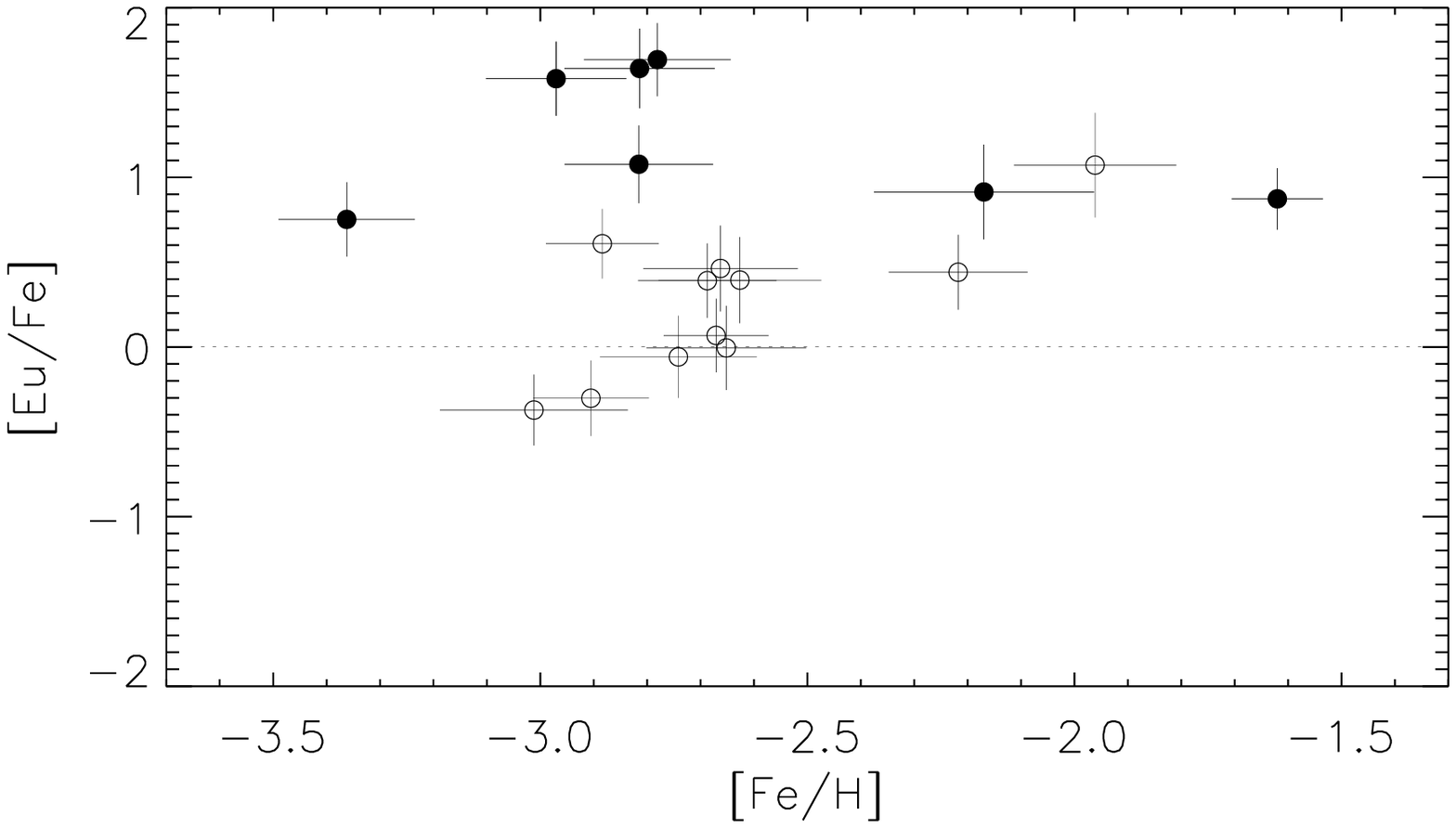}
\caption{The [Eu/Fe] vs. [Fe/H] trend exhibited for stars
in the HERES Pilot Program. Two of the stars with the highest values
of [Eu/Fe] were previously known.}  
\end{minipage}
\end{figure}

\section{HERES Results To Date}

Not all of the HERES spectra obtained thus far have been subjected to detailed
analysis, but this is progressing rapidly. A pilot sample study of the first 35
stars obtained in the HERES program are described in detail in \cite{Bar04}. In
this sample one new $r$-I star, and two new $r$-II stars have been discovered.
This rate of detection of $r$-process-enhanced stars is in line with
expectation, as noted above. In one of the newly discovered $r$-II stars, data
in hand suggest that U is detectable, and new, much higher-quality spectra of
this star have already been obtained; Hill et al. (in preparation) will provide
a detailed analysis. Further inspection of the HERES sample stars, including
data obtained quite recently, has identified many (possibly 7) additional $r$-II
stars, a similar number of $r$-I stars, some 15 $s-$processed enhanced stars,
and other interesting targets for detailed inspection at higher signal-to-noise
ratios.
        
\section {Future Survey Efforts}

Our hope and expectation is that, in the near future, application of a similar
approach to even larger numbers of targets will expand of these classes further,
and/or lead to the identification of new classes of $r$-process-enhanced MP
stars.  Discussion are now underway, for example, to couple medium-resolution
spectroscopic surveys (such as the proposed extension to the Sloan Digital Sky
Survey, SEGUE: Sloan Extension for Galactic Understanding and Evolution) to
high-resolution follow-up with the Hobby-Eberly Telescope (HET).  It is expected that
SEGUE will discover some 10,000 giants with [Fe/H] $< -2.5$, at least a subset
of which will be bright enough for the required high-resolution follow-up to
commence.  While waiting for SEGUE, we are hoping to conduct a exploratory HET
program, using brighter (northern-hemisphere) HK-survey candidates.

\end{document}